**SWSC**

RESEARCH ARTICLE

OPEN ⭗ ACCESS

# Synoptic radio observations as proxies for upper atmosphere modelling

Thierry Dudok de Wit[1,*], Sean Bruinsma[2], and Kiyoto Shibasaki[3]

[1] LPC2E, CNRS and University of Orléans, 3A avenue de la Recherche Scientifique, 45071 Orléans Cedex 2, France
  *e-mail: ddwit@cnrs-orleans.fr
[2] Department of Terrestrial and Planetary Geodesy, CNES, 18 avenue E. Belin, 31401 Toulouse Cedex, France
[3] Nobeyama Solar Radio Observatory/NAOJ, Nagano 384-1305, Japan

## ABSTRACT

The specification of the upper atmosphere strongly relies on solar proxies that can properly reproduce the solar energetic input in the UV. Whilst the microwave flux at 10.7 cm (also called F10.7 index) has been routinely used as a solar proxy, we show that the radio flux at other wavelengths provides valuable complementary information that enhances their value for upper atmospheric modelling. We merged daily observations from various observatories into a single homogeneous data set of fluxes at wavelengths of 30, 15, 10.7, 8 and 3.2 cm, spanning from 1957 to today. Using blind source separation (BSS), we show that their rotational modulation contains three contributions, which can be interpreted in terms of thermal bremsstrahlung and gyro-resonance emissions. The latter account for 90% of the rotational variability in the F10.7 index. Most solar proxies, such as the MgII index, are remarkably well reconstructed by simple linear combination of radio fluxes at various wavelengths. The flux at 30 cm stands out as an excellent proxy and is better suited than the F10.7 index for the modelling the thermosphere-ionosphere system, most probably because it receives a stronger contribution from thermal bremsstrahlung. This better performance is illustrated here through comparison between the observed thermospheric density, and reconstructions by the Drag Temperature Model.

## 1. Introduction

Solar photons in UV wavelengths are the primary heat source of the terrestrial upper atmosphere, which expands and contracts in response to solar activity (Pap et al. 2004). The specification of the upper atmosphere strongly relies on our knowledge of the solar spectrum in the UV, typically between 1 and 300 nm (Floyd et al. 2002; Lilensten et al. 2008; Ermolli et al. 2013). Continuous measurements of the Solar Spectral Irradiance (SSI) at these wavelengths really started in 2003 only, with the launch of the SORCE satellite. The paucity of observations, the short lifetime of the instruments and their difficult long-term calibration have turned the making of continuous SSI records into a major challenge.

In contrast, most users of solar inputs, such as atmosphere and climate modellers, require long, uninterrupted and well-calibrated records. For that reason, many still prefer solar proxies as alternatives to the true SSI. Foremost among these proxies is the solar radio flux at 10.7 cm, also known as the F10.7 index, which reproduces most of the variability of the UV band and has been continuously measured since 1947 (Tapping 2013). The Sun has also been monitored at a few other centimetric (i.e. microwave) wavelengths but these have received considerably less interest, except for flare studies.

In the centimetric range, most of the radio flux consists of thermal bremsstrahlung and of gyroemissions. Although the microphysics of these processes is well understood, the connection between their intensities and the SSI is a complex one

(White 1999). Attempts have been made to use the potential of multiwavelength radio observations for reconstructing specific bands of the SSI (e.g. Schmahl & Kundu 1995), but applications have been unusually slow to follow.

The main objective of this study is to bring these synoptic radio observations in the limelight by showing that they provide considerable added value to the observation of one single wavelength. We concentrate here on daily observations in the centimetric range, ignoring transients such as flares. Long-term trends have become a topic of great interest (Tapping & Valdés 2011), but are also beyond the scope of this study. We shall instead consider the last three decades and determine how the variability of solar radio emission is related to that of the SSI, and to various solar proxies. To do so, we shall rely on a gapless data set of 56 years of daily observations from the Ottawa, Penticton, Toyokawa and Nobeyama observatories, which has been built for that purpose.

This article is organised in four distinct parts. In Section 2, we discuss the physical emission processes of centimetric radio emissions, present our composite data set and discuss some of its properties. In Section 3, we use a powerful statistical method called Blind Source Separation (BSS) to decompose these observations into different contributions, and show how these are associated with physical processes. Section 4 addresses the practical problem of reconstructing the SSI from synoptic radio observations, while Section 5 considers a concrete application, with the reconstruction of the thermospheric density for satellite drag specification. Conclusions follow in Section 6.

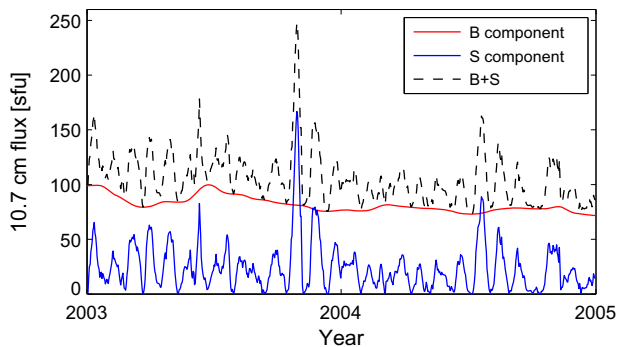

**Figure 1.** Example of the daily-averaged 10.7 cm flux, showing the B and S components, and their sum, which equals the observed flux.

## 2. Synoptic observations: physical properties and results

The Sun-integrated microwave flux is a complex mix of contributions that are governed by various physical processes, which are acting on different regions of the solar atmosphere (Kundu 1965; White 1999; Shibasaki et al. 2011). All of them, however, are eventually structured by the solar magnetic field, as are the processes that produce optical emissions. This common origin explains why microwave radio emissions are so widely considered as tracers of solar magnetic activity, and, by extension, of the variability of the SSI.

Solar radio emissions are traditionally divided into three components (Tanaka & Kakinuma 1958; Kundu 1965), two of which are illustrated in Figure 1:

1. A lower envelope of the time series, which evolves on time scales of months to years, and is sometimes called the *B component*, or background component. This includes a steady background emission that is often called quiet component.
2. A more rapidly varying component that is widely known as the *S component* or rotationally modulated component.
3. On top of these, a highly transient flaring component, which is frequently used to diagnose energetic particle acceleration processes (Pick & Vilmer 2008), but will not be considered further, for the focus on time scales of days and beyond.

The variability of the *S* component results from the combined effect of variability of the radio sources that are located in the solar atmosphere, and the changing transmittivity of the hot solar corona (Tapping & Detracey 1990; White 1999; Shibasaki et al. 2011). Free-free interactions between thermal electrons and ions are one of the important sources of radio emissions. Thermal bremsstrahlung permanently occurs all over the solar disc, but is enhanced by denser plasma trapped in magnetic fields. Such so-called complexes of activity are found in the vicinity of active regions, but may also occur near plages, remnants of decaying active regions, etc. For that reason, they mostly consist of optically thick emissions coming from below the transition region, and from optically thick or thin emissions from dense coronal plasmas. When the magnetic field is strong enough, typically above sunspots, a second, spatially more compact, and much brighter contribution comes from gyroresonance emissions. The latter usually peak at intermediate wavelengths. Indeed, long wavelengths are screened by optically thick bremsstrahlung from the overlying plasma, whereas the maximum intensity of solar magnetic fields sets a lower limit on the wavelength range for gyroresonance emissions. The complexity of the emission process, with emissions occurring at a couple of harmonics, leads to considerable spectral variability with solar activity. For thermal bremsstrahlung, the plasma emissivity is only weakly wavelength-dependent but the observed magnitude is highly sensitive to the opacity of the atmosphere above the source, which also varies with solar activity.

An important open question, which we shall come back to, is the average intensity of thermal bremsstrahlung relative to that of gyroresonance emissions. Schmahl & Kundu (1998) argue that active regions are dominated by gyroresonance emission, whereas Tapping and Detracey (1990) conclude that free-free emissions are prevalent. The difficulty in separating the two mostly stems from the absorption of emissions coming from lower layers by overlying structures such as plages. There is growing evidence, however, for radio emissions at sunspots to be dominated by gyroresonance emissions (e.g. Nindos et al. 2000).

For our purposes, and as a first approximation, we may thus consider the flareless *S*-component in the centimetric range (and in that range *only*) as a linear combination of two contributions with different characteristic spectra: a free-free contribution, which is a tracer of hot and dense regions in the lower corona and below, and a gyroresonance contribution, which is a tracer of sunspots in the photosphere. As it turns out, at 10.7 cm, the mix of the two properly matches the variability that is observed in the Extreme-UV (EUV, 10–120 nm) range. If, however, we are interested in a tracer for the visible range, then radio emissions in the 1–3 cm range are better suited. Reconstructions of the total solar irradiance have indeed be obtained that way (Schmahl & Kundu 1995). Conversely, we expect radio emissions in the 20–30 cm range to be more appropriate for describing the contribution coming from bright features such as plages and faculae, which are highly visible in the UV, and are important for specifying the solar forcing of the thermosphere-ionosphere system. This is precisely what motivates us to investigate the radio flux at various wavelengths, and in particular long ward of 10.7 cm. To substantiate our search for better solar proxies, we shall show in Section 3 that synoptic radio observations can indeed be decomposed by statistical analysis into three independent contributions, which carry the signature of free-free, and gyroresonance emissions.

### 2.1. The data and their properties

Although synoptic radio observations have been made at various observatories, we concentrate here on the Ottawa/Penticton and Toyokawa/Nobeyama facilities, which provide some of the longest, most stable and almost uninterrupted records. Continuity and long history are indeed crucial when it comes to elaborating proxies.

Observations of the 10.7 cm flux have been made routinely by radio telescopes at Ottawa from 14 February 1947 until 31 May 1991, and thereafter by a similar set of instruments at Penticton. Both records are merged into one single time series after correction for antenna gain, etc. The Dominion Radio Astrophysical Observatory provides the daily average flux in near real-time[1]. Adjusted values are devoid of flare signatures and are corrected for a fixed Sun-Earth distance of 1 AU. We multiply all values by 0.9, giving so-called *URSI D values*, which are more consistent with the other wavelengths. For more details, see Tapping (2013).

**Table 1.** The data sets used in this study and their main characteristics. The noise level is estimated from a red noise model. The number of gaps refers to the number of missing days in the online data files, not to the number of actual service interruptions, which is smaller. Averages refer to fluxes adjusted at 1 AU.

| Wavelength (cm) | Frequency (GHz) | Origin of observations | Beginning of measurements | Number of gaps since beginning | Average level (sfu) | Noise level (sfu) |
|---|---|---|---|---|---|---|
| 3.2 | 9.4 | Toyokawa/Nobeyama | May 1, 1956 | 195 | 297 | 3.0 (1.0%) |
| 8.0 | 3.75 | Toyokawa/Nobeyama | Nov. 6, 1951 | 203 | 124 | 2.5 (2.0%) |
| 10.7 | 2.8 | Ottawa/Penticton | Feb. 14, 1947 | 311 | 111 | 2.8 (2.5%) |
| 15.0 | 2.0 | Toyokawa/Nobeyama | June 1, 1957 | 231 | 101 | 2.0 (2.0%) |
| 30.0 | 1.0 | Toyokawa/Nobeyama | March 1, 1957 | 163 | 81 | 1.9 (2.4%) |

The Toyokawa/Nobeyama radio polarimetres monitor the Sun in multiple frequencies. Here we consider observations made at 30 cm, 15 cm, 8 cm and 3.2 cm only. Shorter wavelengths are observed as well, but we ignore them because of their increasing sensitivity to weather conditions. Observations began on 6 November 1951 in Toyokawa at 8 cm, see Table 1. From 24 February 1994 to 14 May 1994, all but the observations at 8 cm were interrupted as the antennas were moved from their location at Toyokawa to nearby Nobeyama. All instruments operate continuously during daytime. Once per month, flare-corrected daily averages are published[2]. Lower-level daily averages are also automatically published on a daily basis. These fluxes are not normalised to a constant Sun-Earth distance. A more detailed description can be found in Tanaka et al. (1973).

Although ground-based radio instruments suffer far less from the degradation and outages that plague space-borne optical instruments, their absolute calibration remains an important and challenging issue (Tanaka et al. 1973; Tapping & Charrois 1994). All instruments mentioned above are routinely calibrated at least once per day, always using the same procedure, which guarantees the consistency of the observations on the long term. This continuity is a unique asset of these radio observations.

The instrumental error on the observations is estimated to be of the order of 1%. We obtain values between 1 and 2.5% when using a different estimator, which assumes a red noise spectrum (Mann & Lee 1996), see Table 1. To simplify, in what follows, we shall assume the relative error to be 2.5% for all wavelengths. This value is indicative, in the sense that it does not allow for distinguishing between different types of noise, and does not properly account for the heteroscedasticity of the noise.

Figure 2 shows the raw daily values at all five wavelengths, with a conspicuous 11-year solar cycle modulation. Note that the base level reached at each solar minimum is barely cycle-dependent. We shall not address this issue, which is of great interest, but beyond the scope of our study.

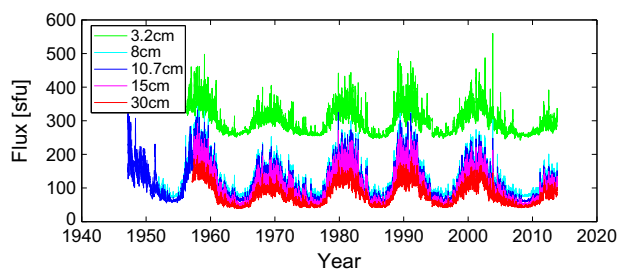

**Figure 2.** Time evolution of the original data, normalised to 1 AU, at all five wavelengths. All quantities are expressed in solar flux units (sfu).

## 2.2. The making of a single composite

An important prerequisite for analysing solar variability is to start with a homogeneous data set that has a common time grid for all observations, and no data gaps. For that reason, we first stitch together all observations into a single composite data set. Here, we consider daily averages at five wavelengths: 3.2, 8, 10.7, 15 and 30 cm. The time span is set by the interval over which all five wavelengths are monitored, and ranges from 6 November 1957 to 31 July 2013. The number of flux measurements made each day, and their times vary between observatories. However, the primary measurement is generally made at local noon: 17:00 UT in Ottawa, 20:00 UT in Penticton and 3:00 UT in Nobeyama and Toyokawa. For that reason we regrid them all to the same 12:00 UT time stamp by using cubic splines.

The number of data gaps for our interval is given in Table 1. Most gaps are one or two days long, except for the 79-day discontinuation between Toyokawa and Nobeyama. During that period, however, the 8 cm and the 10.7 cm continued almost uninterrupted. Since all five observations are highly correlated between each other, and are also correlated in time, we have an excellent example wherein data gaps can be filled by expectation-maximisation. This approach consists in reconstructing the missing values by interpolation in time and across wavelengths, assuming that the values are coherent in time and across wavelength (see Dudok de Wit 2011). The error on the reconstructed values is approximately 2%, which is comparable to the instrumental error given in Table 1, and thus confirms the excellent performance of the gap filling procedure. Filling in all these data gaps is not mandatory for our analysis but certainly eases it, especially when it comes to estimating quantities such as the power spectral density.

---

[1] The data are available at ftp://ftp.ngdc.noaa.gov/STP/space-weather/, in the directory /solar-data/solar-features/solar-radio/noon-time-flux/penticton/

[2] These data are available at http://solar.nro.nao.ac.jp/norp/html/daily_flux.html

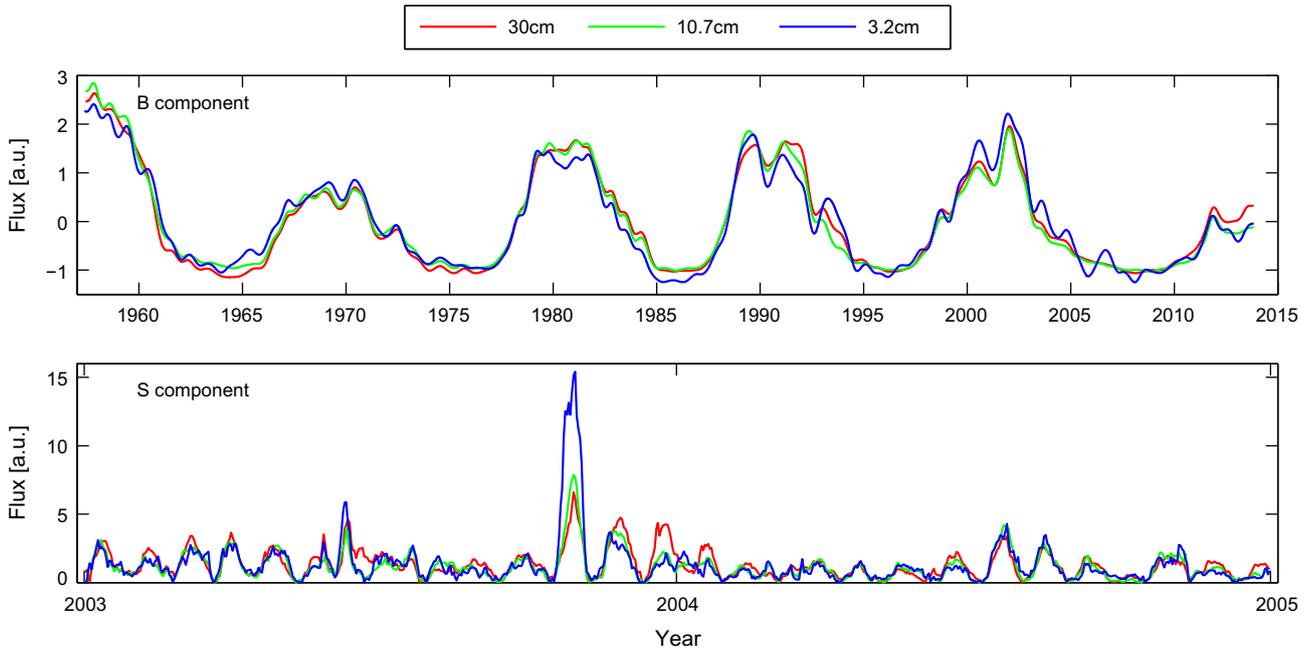

**Figure 3.** Excerpt of the flux at three wavelengths with the background component, averaged over 6 months (top), and the *S* component (bottom). All values have been rescaled, and baselines have been shifted vertically to emphasise their tiny differences. Arbitrary units have been used.

Finally, we normalise all values to a constant Sun-Earth distance of 1 AU and by convention express them in solar flux units, with 1 sfu = $10^{-22}$ W m$^{-2}$ Hz$^{-1}$. The total number of samples is 20 357. This data set is available for download from http://projects.pmodwrc.ch/solid/.

To separate the variability into *B* and *S* components, we use the same procedure as Dudok de Wit & Bruinsma (2011): the envelope is extracted by taking the minimum value in a sliding 21-day window and subsequently lowpass filtered with a cutoff at 21 days. We found this number of days to provide the best separation between the rotational variation and the slowly changing radio flux. We obtain qualitatively similar results by running a lowpass filter only, but the first procedure offers some advantages. First, as discussed by Schmahl & Kundu (1998), there is a physical reason for considering the lower level of activity for the background component as it is robust to bursts of activity. Second, this procedure yields a non-negative *S* component, which can thus be interpreted as a physical flux. This separation is illustrated in Figure 1.

### 2.3. Observed variability

Figure 3 gives an overview of radio data. The background component exhibits a remarkably similar variability at all wavelengths, with small differences that barely exceed the measurement error. The *S* component is equally coherent, with more marked differences. Note in particular how the intense active regions during the Halloween events of October and November 2003 cause a peak of intensity at short wavelengths whereas in the months that follow, the remnants of these active regions lead to an excess of flux at long wavelengths.

To better understand the differences between the various wavelengths, we show in Figure 4 two scatter plots with the fluxes at 3.2 and 30 cm versus the 10.7 cm flux. We point out that background components have different offsets (i.e. different quiet Sun levels) and that their relationship is almost linear. From this we conclude that proxies based on the background component only are almost interchangeable, provided that their offset is taken into account.

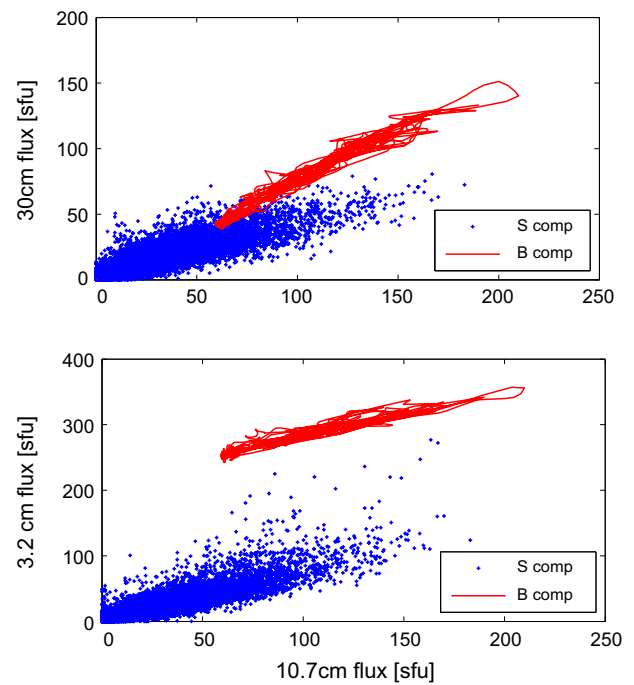

**Figure 4.** Scatter plot of the 30 cm flux (top) and the 3.2 cm flux (bottom), versus the 10.7 cm flux. The *B* and *S* components are represented separately.

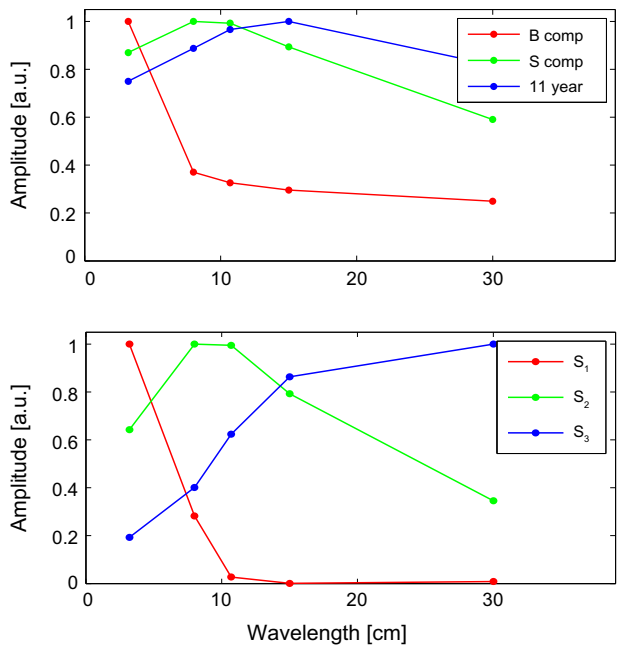

**Figure 5.** Mean amplitude of the $B$ and $S$ components, and amplitude of the 11-year solar cycle modulation observed in the background component (top), and mean amplitude of the three source terms discussed in Section 3 (bottom). All quantities have been normalised to their maximum value in order to ease visualisation.

The second important result is the scatter in the $S$ components, which cannot be ascribed to noise only, but is also due to differing dynamics, see Figure 3. Note that the slope of the cloud of points differs from that of the $B$ component. Therefore, the solar cycle modulation and the solar rotational modulation have distinct signatures in the radio spectrum, which implies that none of the fluxes can be expressed as a static linear or nonlinear function of the others. Identical conclusions had already been reached by Dudok de Wit & Bruinsma (2011) for EUV proxies.

Figure 5a illustrates another aspect of the variability, by comparing the mean amplitudes of the $B$ and $S$ components, as well as the modulation amplitude of the 11-year solar cycle. The peak we observe in the S component near 10 cm has been interpreted by Schmahl & Kundu (1995) as an evidence for the prevalence of gyroresonance emissions, whereas Tapping & Detracey (1990) consider it as the signature of thermal free-free emissions. We shall provide new evidence for this issue in Section 3.

Most solar signals exhibit a strong 27-day fundamental period with a couple of harmonics. These harmonics mainly stem from the centre-to-limb variation of the emission processes (Donnelly & Puga 1990), and their periodic occultation. Up to two harmonics can be observed in the radio fluxes, see Figure 6, but their amplitudes are considerably larger at 15 and 30 cm. This difference cannot be reasonably explained in terms of centre-to-limb variations only; the most plausible explanation is to be found in the lifetime of the solar features that contribute to the radio flux. Indeed, there are features producing bremsstrahlung which may last longer than active regions, which are the primary source of gyroresonance emissions. For that reason, we expect bremsstrahlung to exhibit a more coherent modulation by solar rotation, which then offers a better chance for harmonics to build up coherently. Stronger harmonics at long wavelengths thus are an indication for a greater contribution from bremsstrahlung.

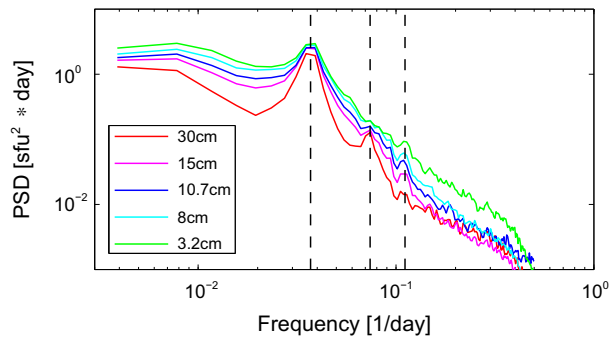

**Figure 6.** Power spectral density of the fluxes, after standardising the original data. The dashed lines indicate the position of the 27-day fundamental frequency, and its first two harmonics. These spectra were estimated by windowed Fourier transform.

## 3. Separation into different contributions

Since our observations consist of a mix of constituents with different spectral signatures in their UV counterpart, it would be interesting to actually extract these constituents. This inverse problem arises in various other disciplines wherein one has a mixture and would like to recover from it the original constituents, called sources. The problem is called *blind source separation* because both the sources and their concentrations have to be inferred, using the least prior knowledge (Comon & Jutten 2010). Many techniques have been developed for that purpose. The most elaborate ones rely on Bayesian approaches, which have been applied with success in chemometrics (Moussaoui et al. 2006), in astrophysics (Kuruoglu 2010) and to solar irradiance (Amblard et al. 2008; Dudok de Wit et al. 2012). From now on, the word *source* will systematically refer to one of these elementary constituents of the solar radio fluxes, as determined by statistical means, and not to the physical sources of radio emissions.

Since our BSS problem is ill-posed, it is essential to incorporate the proper constraints in the method. Here, we assume that the integrated radio flux at each wavelength is a linear and instantaneous combination of emissions coming from different sources, all of which are positive (no negative emissions) and with positive concentrations only (emissions can only add up, not interfere negatively). In addition to that, we seek sources whose time evolutions are mutually independent in a probabilistic sense. We consider here the Bayesian positive source separation technique by Moussaoui et al. (2006), which assumes that each flux $\Phi[\lambda, t]$ can be decomposed as:

$$\Phi[\lambda, t] = \sum_{k=1}^{N_S} A_k[t] S_k[\lambda] + N[\lambda, t], \qquad (1)$$

where $A[t] \geq 0$ are the time-dependent concentrations, $S[\lambda] \geq 0$ are the spectral profiles of the sources and $N[\lambda, t] \geq 0$ is a noise term that encompasses measurement errors and model uncertainties; here we assume a gamma distribution, which is flexible enough to encode various types of noise distributions. In the Bayesian framework, $A$ and $S$ are random matrices whose posterior probability distribution is computed. More details can be found in Moussaoui et al. (2006). The uniqueness and stability of the solution is an important issue in BSS. We tested other techniques such as non-negative matrix factorisation (Lee & Seung 1999), and the overall excellent agreement between various techniques lends strong support to the analysis that follows.

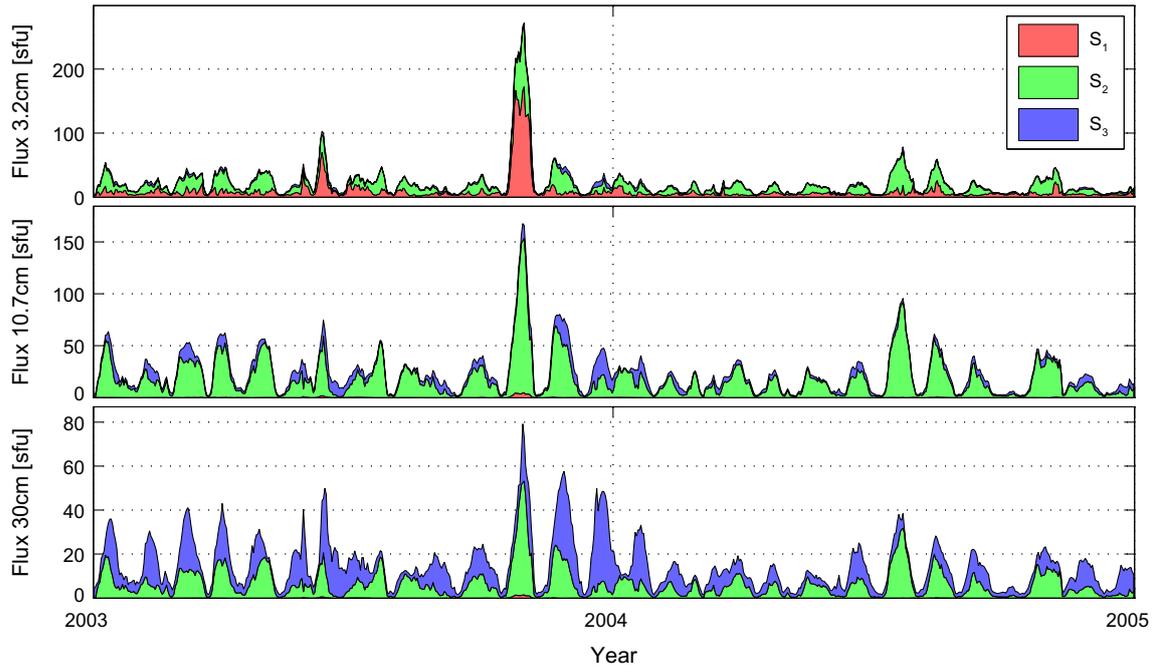

**Figure 7.** Decomposition of the *S* component into its three sources, as estimated by positive BSS.

The main free parameter in all BSS methods is the number $N_S$ of sources, which is usually bounded by the number of variables, here five. When decomposing the data into 1–5 sources, and subsequently comparing the reconstruction $\hat{\Phi}$ to the original flux $\Phi$, we obtain average root mean squared errors (RMS) of respectively 4.9, 3.7, 2.3, 2.2 and 2.1 (sfu). We use the classical definition of the RMS:

$$\varepsilon = \sqrt{\left\langle \left(\Phi - \hat{\Phi}\right)^2 \right\rangle_t}, \qquad (2)$$

where $\langle \ldots \rangle_t$ stands for time-averaging. The RMS gradually decreases when the number of sources increases from one to three but saturates thereafter around a value that is comparable to the noise level (see Table 1), which suggests that the fit does not improve further. Other tests, such as the reproducibility of the results, also suggest that $N_S = 3$ provides a good compromise between error reduction and model complexity. We shall thus keep that number in what follows. The sources we obtain are robust and unique in the sense that they are barely affected when the analysis is restricted to specific solar cycles only. The Bayesian approach gives access to their posterior probability distribution, whose standard deviation is always of the order of, or less than 1%, and can thus be neglected.

The source profiles $S[\lambda]$ we obtain by BSS are shown in Figure 5b. From the discussion in Section 2, we can tentatively interpret source 2 as a gyroresonance contribution, because it peaks at intermediate wavelengths, while source 3 complies more with the spectral response of thermal bremsstrahlung. Let us next consider the time-dependent concentration of the sources, see Figure 7; we display an excerpt of the fluxes at 3.2, 10.7 and 30 cm, after prior decomposition into their three contributions. Source one only contributes during periods of major complex sunspots with exceptionally intense magnetic fields, which supports a connection with gyroresonance processes. Source 3 is omnipresent and is most intense in the aftermath of major active regions, during times when bright remnants cover the solar surface. This, together with its spectral profile, pleads for a strong contribution from thermal bremsstrahlung. Source 2 had been ascribed to gyroemissions, and this is further supported by its enhancement during times with sunspots, but not necessarily when there are more plages or faculae.

### 3.1. Comparison with solar proxies

To substantiate our interpretation, let us now consider the *S* component of various common solar proxies, using exactly the same procedure as above. One could repeat the procedure by extracting again their sources by BSS. However, we are primarily interested in determining what fraction of the proxies can be explained in terms the three sources of the radio fluxes. To this end, we model the rotational modulation $y[\lambda, t]$ of each proxy as a linear combination of these sources:

$$y[\lambda, t] = \sum_{k=1}^{3} \alpha_k A_k[t] + \varepsilon[\lambda, t]. \qquad (3)$$

The residual error $\varepsilon$ represents the fraction of the observations that is not correctly fitted by the model. As a measure of the relative contribution each source, we consider the standard deviation of each of the terms on the right hand side of equation (3) and divide them by their sum. Also included in this sum is the residual error. The model is fitted only for the time interval for which each proxy is available. We consider the following set of common solar proxies:

- ISN: the international sunspot number (from the Royal Observatory of Belgium).
- DSA: the daily sunspot area (Royal Greenwich Observatory).

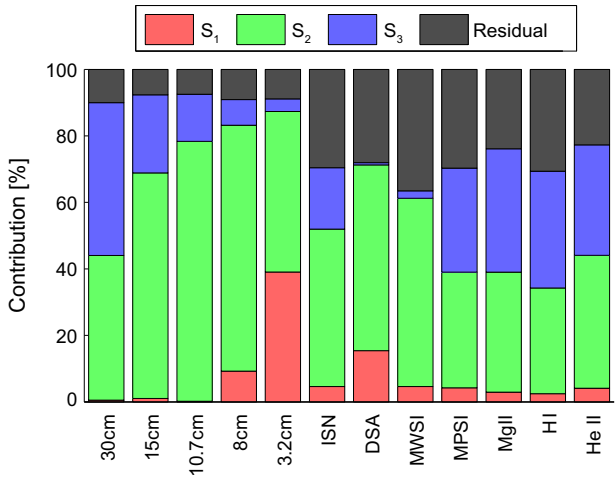

**Figure 8.** Relative contribution of each of the three sources to the five radio fluxes and to seven common solar proxies. For each quantity, the sum of the contributions has been normalised to 100%.

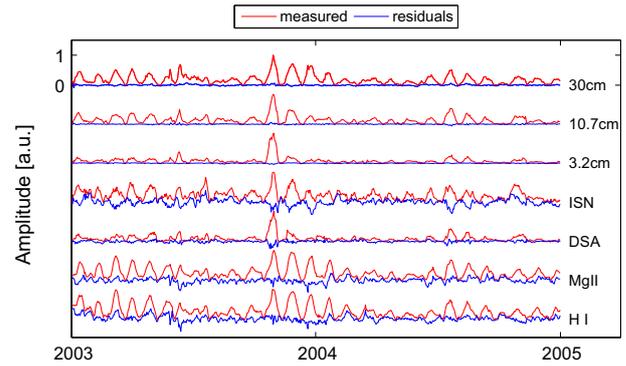

**Figure 9.** Excerpt of the residuals, shown for a subset of the radio fluxes and solar proxies. The observations are in red, and the residuals in blue, with the same units. For ease of visualisation, a different scale has been used for each proxy.

- **MWSI:** the Mount Wilson Sunspot Index (Mount Wilson Observatory), which is derived from the fractional solar surface covered by pixels for which the magnetic field exceeds 100 Gauss. As its name suggests, this is a proxy for sunspots.
- **MPSI:** the Magnetic Plage Strength Index (Mount Wilson Observatory), which is derived from the fractional solar surface covered by pixels for which the magnetic field is between 10 and 100 Gauss. This is a proxy for plages and faculae.
- **MgII:** the core-to-wing ratio of the Magnesium K line (LASP composite), which is a proxy for the solar UV irradiance.
- **H I:** the intensity of the H Lyman-$\alpha$ line at 121.57 nm (LASP composite), which is the most intense line in the UV band.
- **He II:** the irradiance in the 26–34 nm band (SEM, onboard SoHO), which is dominated by the bright chromospheric He II line at 30.4 nm.

Figure 8 summarises the results and confirms the excellent reconstruction of the radio fluxes by the three sources. The residual error is several times larger for the proxies than for the radio data, but its value remains reasonable. Not surprisingly, source one appears almost exclusively in proxies that describe highly active regions, such as the DSA. Source 3, on the contrary, dominates in the MPSI and in the UV lines, which are sensitive to the occurrence of plages, faculae and bright loops. The attribution of source 2 to gyroresonance emissions is consistent with this picture.

The residuals associated with each reconstruction are illustrated in Figure 9, for some of the quantities. The largest residuals are observed with the sunspot number, presumably because that proxy is more loosely connected to the radiative output of the Sun.

Let us stress that our three sources are unlikely to capture one single physical process, even though their characteristics leave little doubt about the prevalent contribution. Source 3, for example, may receive contributions coming from coronal loops, in addition to plages and faculae. Source 1 is the most puzzling one, as it peaks at short wavelengths, and is only occasionally active. Our analyses suggest that it has at least two contributions. One is the sporadic pollution by flares, which are not always properly removed from the daily value of the flux. Flares, for example, are responsible for the small peak observed in source 1 on 10 June 2003, see Figure 7. A second and more important component is associated with complex sunspots that have intense magnetic fields. For such events, the gyro-resonance layers of the 3.2 and 8 cm fluxes are located high in the corona, which leads to enhanced emissions at those short wavelengths. Incidentally, this increase in short wavelength emissions is known to be highly correlated with the occurrence of proton flares (Tanaka & Enome 1975), which opens the perspective of using source 1 as a proxy for such flares.

In agreement with Schmahl & Kundu (1998) and others, we now conclude that the main contribution to the rotational modulation of the radio flux comes from gyroresonance emissions, rather than from bremsstrahlung. To the best of our knowledge, this is the first example wherein these individual contributions have been separated without the use of spatial information. For emissions at 10.7 cm, we find that free-free emissions account for about 10% of the variability, whereas at 30 cm, their share increases to about 50%.

The same BSS analysis could in principle be applied to the background component. However, the strong coherency of the variability does not give enough leverage for distinguishing its sources. Furthermore, our Bayesian technique yields large confidence intervals for the sources, which excludes any meaningful interpretation. The default model for the background component thus consists of one single source with a variable offset.

## 4. Reconstruction of the spectral irradiance and of solar proxies

A key issue regarding applications for space weather and space climate is the possibility of using synoptic radio observations (or their sources) for reconstructing specific bands of the SSI. This raises two practical questions: Are the sources we obtain by BSS better suited for reconstructing specific spectral bands or should we use instead combinations of wavelengths? In the latter case, what are those best wavelengths?

The first question can be readily answered by noting that our BSS sources are not ideally suited for practical and operational applications. Indeed, the $S$ component is not robust to outliers, requires a non-causal filter, and the extraction of the sources is computationally expensive. Nevertheless, because of high relevance of the sources for their physical interpretation, let us still compare them to the SSI.

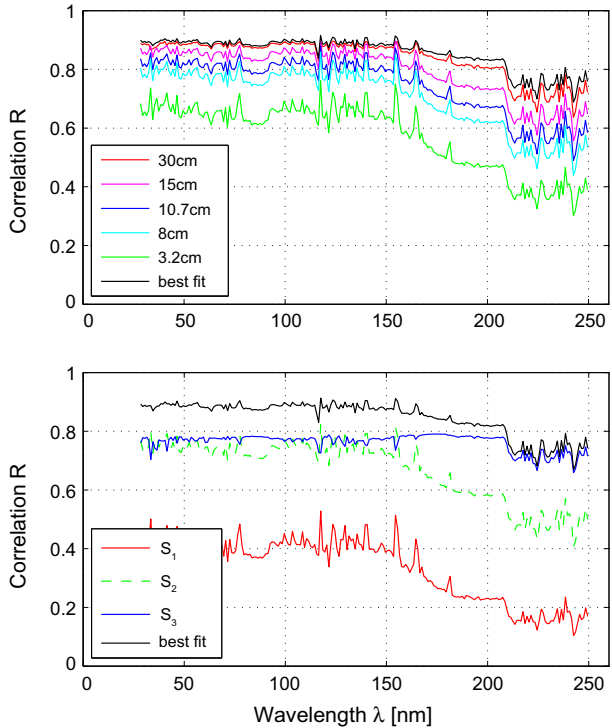

**Figure 10.** Correlation coefficient between the SSI and the radio flux at various wavelengths (top), and with the three sources (bottom). Only the variable parts ($S$ component) are compared. Best fit refers to the correlation coefficient obtained with the linear combination of wavelengths, respectively sources, that best matches the SSI.

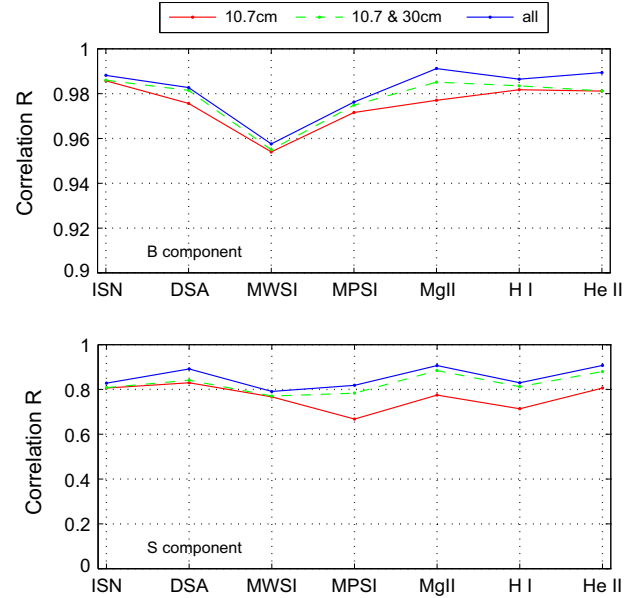

**Figure 11.** Correlation coefficient between the radio flux at various wavelengths, and nine proxies. The upper plot is for long time scales only, and the lower one for short time scales. Note that the vertical scales differ.

Our SSI observations span the period running from 15 May 2003 to 31 July 2013, during which the entire UV spectrum was continuously observed. This data set is a composite of observations made by the TIMED/SEE instrument (Woods et al. 2005) for wavelengths ranging from 28 to 115 nm (version 11), and by the SORCE/SOLSTICE instrument (McClintock et al. 2005) for wavelengths from 115 to 250 nm (version 12). We discard wavelengths outside of this range either because they suffer too much from instrumental artefacts, or because the estimation of the S component is compromised, or both. Here, we consider Pearson's correlation coefficient $R$ as a means for comparing the each source to the SSI; a squared value of $R^2 = 0.95$ means that the source explains 95% of the variance seen in the SSI.

Figure 10 summarises the results and shows that all five radio wavelengths correlate reasonably well with the SSI. However, the highest correlations are systematically found with the longest radio wavelengths, which therefore stand out as the best proxies for the rotationally modulated part of the SSI. The gain in correlation versus the usual 10.7 cm flux ranges from 0.08 to 0.15, which is considerable. Interestingly, the correlation hardly improves further when a combination of wavelengths is used instead (see black curve), which suggests that there is no compelling need for selecting more than one or two wavelengths for reconstructing the SSI.

Figure 10 also reveals the good performance of source 3 for reconstructing the SSI, although the levels usually do not match those obtained with radio fluxes. Source 3 stands out above 160 nm and in the Lyman continuum below 91 nm, that is, in spectral bands associated with emissions coming from the transition region, or below. Source 2 dominates at shorter wavelengths (and in particular below 10 nm, not shown) and for hot coronal lines. The sharp decrease above 210 nm is an instrumental artefact. These results again corroborate the association between source 3 and features such as plages and faculae.

At this stage we conclude that the radio fluxes, rather than their individual sources, generally provide a better reconstruction of the rotational modulation of the SSI because they contain a mix of gyroresonance and bremsstrahlung emission. The best overall performance is obtained with the flux at 30 cm, because it has a stronger bremsstrahlung component. A pure bremsstrahlung proxy (i.e., source 3) is interesting only for wavelengths beyond approximately 180 nm.

The slowly varying baseline could in principle be analysed in the same way, but we shall skip it here, for two reasons. First, as mentioned before, no sources can be extracted from it. Second, several wavelengths of the SSI suffer from degradation and hence do not faithfully reproduce long-term variations. For that reason, great care should be taken in interpreting these values.

The correlation between the radio fluxes and the proxies is summarised in Figure 11, in which our seven proxies are compared to the 10.7 cm flux only, to the best linear combination of the 10.7 and 30 cm fluxes and to the best linear combination of all five fluxes. The relative gain in correlation obtained by combining the 30 cm and 10.7 cm wavelengths is modest, except for proxies that describe the variability in the UV, namely the MPSI, the MgII index, and the H I and He II lines; the correlation typically increases by 0.09–0.13 for the $S$ component. We observe no significant gain with more than two wavelengths. Among the various combinations we tested, the winning combination often was 10.7 cm and 30 cm. In the background component, the average correlation is high from the outset, and the improvement is more modest, though still significant as compared to the confidence interval of $R$.

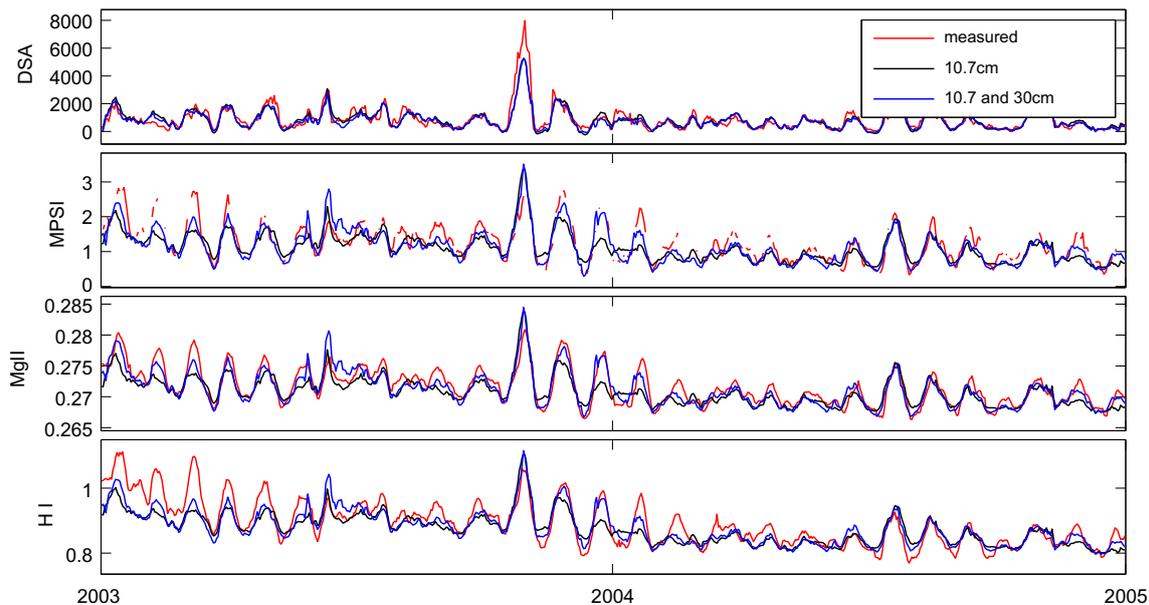

**Figure 12.** Excerpt of four solar proxies, and their reconstruction with the 10.7 cm flux only, and with both the 10.7 and 30 cm fluxes. From top to bottom: the DSA, the MPSI, the MgII index, and the intensity of the H I line in (mW/m²/nm). The time interval is identical to that used in Figure 7.

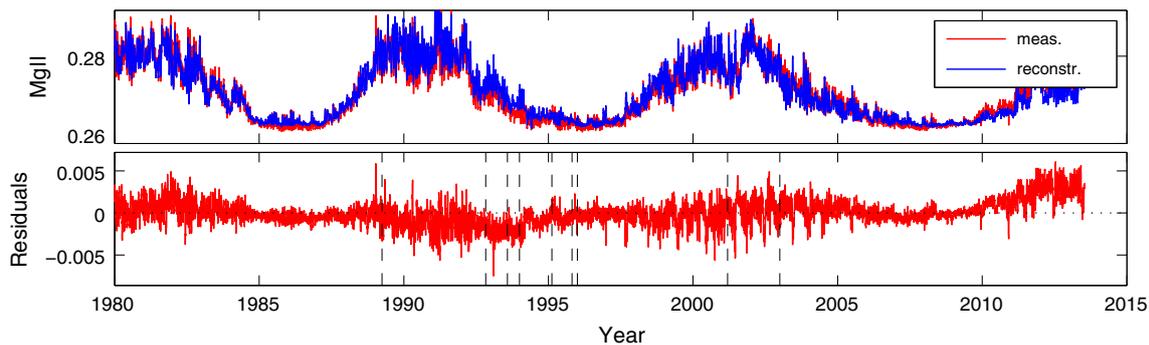

**Figure 13.** Reconstruction of the composite MgII index by using synoptic radio observations at 10.7 cm, and 30 cm. The residual error is expressed in the same (dimensionless) units as the MgII index. Instances at which a different instrument was used to estimate the MgII index are shown by dotted lines. For details on the MgII composite, see Viereck et al. (2001).

To summarise, by combining two or more synoptic radio observations we significantly improve the reconstruction of proxies that describe UV emissions, but less so for proxies such as sunspot numbers. Two wavelengths already provide a good compromise between model complexity and reconstruction. Among all combinations, we recommend the 10.7 cm and 30 cm wavelengths.

To illustrate the reconstruction of proxies, we compare in Figure 12 the reconstruction of four solar proxies by using between one and two wavelengths. Not surprisingly, the quality of the fit is highest for proxies that capture the UV variability of the Sun, such as the MgII index. The improvement brought by the 30 cm flux is particularly obvious in the months that followed the Halloween events of October and November 2003. There are also discrepancies, such as the H I line during early 2003. This mismatch mostly comes from a difference in the evolution of the background component, whose reconstruction is still open for improvement; some discrepancies may also be due to the sensitivity of the H I line to the active network.

Clearly, this is only a first step towards a model that can be improved in several ways. Better reconstructions can be achieved by decomposing all the observations in more than two components, for example by using multiresolution analysis. A more thorough statistical analysis, with the comparison of different metrics, is now at order to compare the various options more in detail.

One interesting spinoff is the assessment of proxy homogeneity. Many proxies are composites that suffer from non stationary noise. Radio data, on the contrary, are highly stable in time. Therefore, if we assume that the connection between the radio fluxes and the proxies does not change in time, then the residuals (i.e. the difference between the observed and the reconstructed proxy) should be stationary as well. Figure 13 shows an application to the MgII index, whose residuals exhibit clear signatures of non-stationarity. Interestingly, several of the discontinuities in the residuals coincide with periods at which a different instrument was used for building the composite. The stability of radio fluxes offers here an opportunity for empirically testing solar proxies and the SSI for outliers, and for anomalous changes.

**Table 2.** The data used in the construction of DTM2012.

| Data | Time frame |
| --- | --- |
| CHAMP | 05/2001–08/2010 |
| GRACE | 01/2003–12/2010 |
| Starlette & Stella | 01/1994–12/2010 |
| Deimos-1 | 03/2010–09/2011 |
| CACTUS | 07/1975–01/1979 |
| OGO6 (T) | 06/1969–08/1975 |
| DE-2 (T, He, O, N2) | 08/1981–02/1983 |
| AE-C (N2) | 01/1974–04/1977 |
| AE-E (T, He, O) | 12/1975–05/1981 |

## 5. Modelling of the thermospheric density

Let us now consider a concrete test case by comparing the 30 cm flux to the routinely used 10.7 cm flux in thermosphere density modelling. From now on we shall follow the convention wherein the latter is called the F10.7 index, and likewise the 30 cm flux will be renamed F30 index.

Thermosphere models are, besides their use in atmospheric studies, necessary to calculate the atmospheric drag force in satellite orbit computation. They predict temperature and composition as a function of the location (altitude, latitude, longitude, local solar time), solar and geomagnetic activities, and day-of-year. The latest version of the Drag Temperature Model, DTM2012 (Bruinsma et al. 2012), uses F10.7 as solar proxy and will be compared below with a version constructed identically but with F30. These models reproduce the climatology of the thermosphere and have a low spatial resolution of the order of thousands of kilometres. All density variations with smaller scales are sources of geophysical noise. The temporal resolution is limited by the solar and geomagnetic activity indices, 1 day and 3 h, respectively. The solar F10.7 emission is used in DTM (and by most models) as a proxy for UV/EUV emissions, which heat the upper atmosphere of the Earth. The JB2008 model (Bowman et al. 2008) additionally uses measurements from space, such as He II and MgII, but they suffer from frequent data outages, inhomogeneous quality, and calibration issues. A ground measurement is in our opinion preferential, and even more so for operational thermosphere models, in order to avoid such issues.

Total neutral densities inferred from high-resolution, high-accuracy accelerometer measurements on the Challenging Mini-Satellite Payload (CHAMP) satellite in the altitude range 450–250 km (Bruinsma et al. 2004), and the Gravity Recovery and Climate Experiment (GRACE) satellite near 490 km altitude were used in the model construction. These two data sets are the main suppliers of the DTM model since the 2009 version. Daily-mean densities inferred from orbit analyses, mass spectrometer data from the Atmosphere Explorer satellites and the Dynamic Explorer 2 satellite (temperature, and O, He and $N_2$ partial densities), as well as OGO6 temperatures were also used. The data sets used in DTM2012 are listed in Table 2. For more details on the historic data sets, the reader is referred to Bruinsma et al. (2003).

The reproduction of total density in the altitude range 120–1500 km is achieved by summing the contributions of the main thermosphere constituents ($N_2$, $O_2$, O, He, H), under the hypothesis of independent static diffuse equilibrium.

The height function $f_i(z)$ per constituent $i$ is the result of the integration of the differential equation of diffusive equilibrium; partial densities specified at 120 km altitude are propagated to higher altitudes employing this function. The exospheric temperature and the partial density variations as a function of the environmental parameters $L$ (latitude, local solar time, solar flux and geomagnetic activity) are modelled by means of a spherical harmonic function $G(L)$. The total density $\rho$ at altitude $z$ is then calculated as follows:

$$\rho(z) = \sum_i \rho_i(120\,\text{km})f_i(z)e^{G_i(L)}. \qquad (4)$$

DTM2012 models the exospheric temperature and the atmospheric constituents each with up to 50 coefficients, which are fitted in a least-squares procedure to the data listed in Table 2. The function $G(L)$ is used to describe periodic and non-periodic variations. Periodic variations are defined as annual and semi-annual terms, as well as diurnal, semidiurnal and terdiurnal terms. The non-periodic terms consist of constant zonal latitude coefficients, and coefficients relating solar and geomagnetic activity to temperature and density. DTM2012 uses F10.7 as solar proxy; for this study, a version was developed using the same density data, but with F30 instead. The models will be referred to as DTM2012_F10 and DTM2012_F30 in the following.

### 5.1. Model evaluation

The models are compared with total density data, both used and not used in the construction. They are evaluated by computing the observed-to-calculated (O/C) density ratios, which stand for the relative precision of the models. The mean of O/C represents the bias whereas the RMS is a combination of the ability of the model to reproduce the observed variations and the geophysical and instrumental noise in the observations. The RMS is defined as in equation (2), except that we use (1) as the expectation for O/C. Pearson's correlation coefficient $R$ between the observed and computed densities is also estimated.

The data used in the evaluation are densities from CHAMP and GRACE, and independent data sets from the US Air Force. Accurate daily-mean densities computed with the EDR (Energy Dissipation Rate) method are used: two spacecraft below 300 km, one at almost 400 km, and five between 400 and 500 km altitude. These were not assimilated in DTM2012. The results are averaged per year, and the EDR densities in the 400–500 km are furthermore averaged over five spacecraft. The mean and RMS of the density ratios are displayed in Figure 14 per year for these data sets. The figures clearly show that DTM2012_F30 is an improvement over DTM2012_F10.

The mean and standard deviation of the annual density ratio time series, the average RMS and correlation coefficients are listed in Table 3. The mean O/C is often closer to 1 with DTM_F30 than with DTM_F10. However, what matters most is the lower dispersion of the mean, the relatively smaller value of the RMS, which decreases by 7%, and the higher correlation when DTM_F30 is used.

A second independent density data source from the US Air Force is the operational model HASDM, or High Accuracy Satellite Drag Model (Storz et al. 2005); this model is updated in near real-time from observed drag effects on about 70 objects using radar tracking data. The densities along the orbit of ESA's Earth Explorer satellite GOCE (Gravity field and steady-state Ocean Circulation Explorer, Drinkwater et al. 2003) at an average altitude of 270 km were provided for the years 2010–2012.

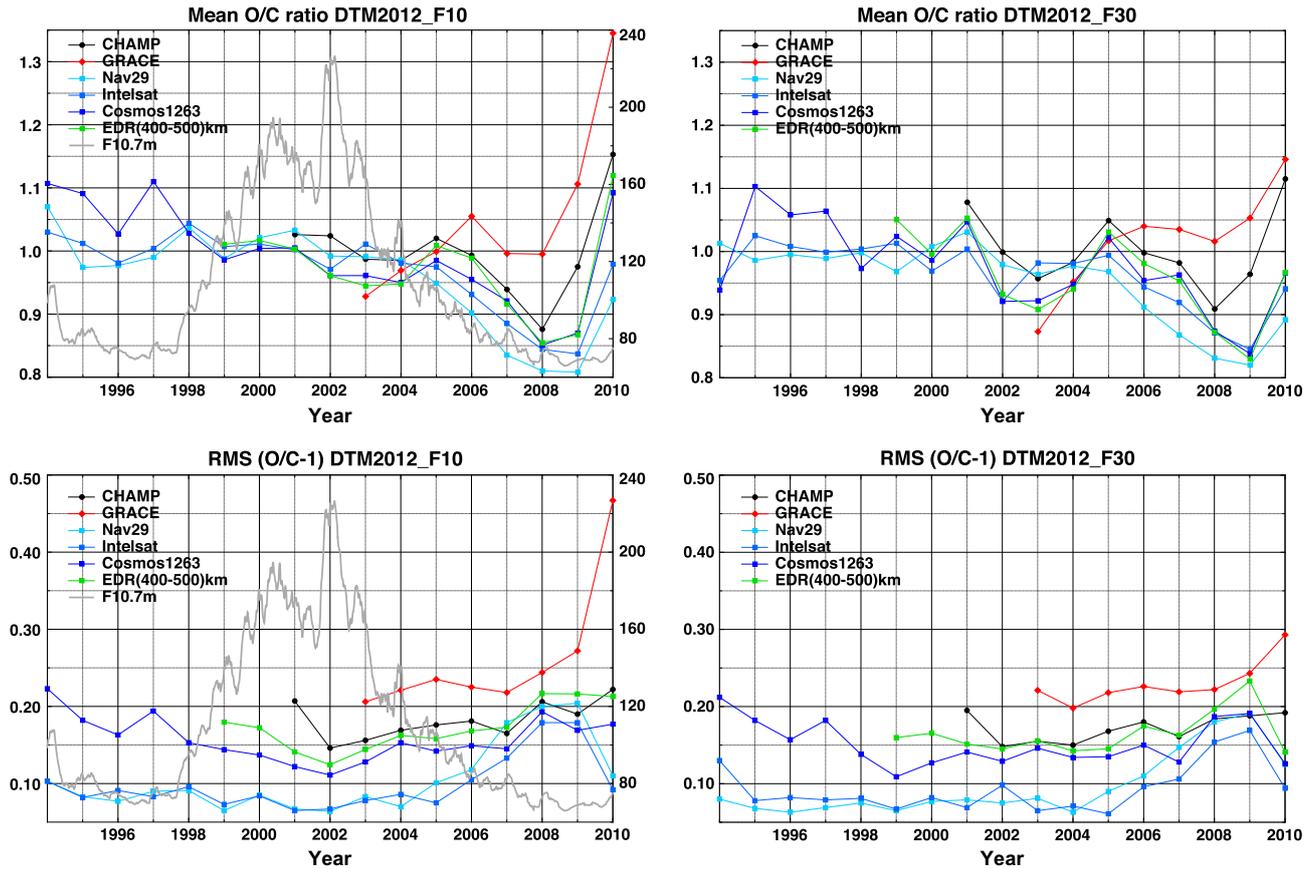

**Figure 14.** Density ratios, per year (upper row) and RMS of density ratios, per year (lower row). These values are computed with DTM2012_F10 (left column) and DTM2012_F30 (right column). The average F10.7 is shown in grey, in arbitrary units.

**Table 3.** Statistics of the model comparisons with CHAMP, GRACE, US Air Force EDR, and HASDM densities. The first of each pair of values stands for the result obtained with DTM2012_F10, and the second one with DTM2012_F30.

|  | Mean *O/C* | $\sigma$ Mean *O/C* | RMS *O/C* | *R* |
|---|---|---|---|---|
| CHAMP | 0.998; 1.003 | 0.070; 0.061 | 0.179; 0.172 | 0.937; 0.941 |
| GRACE | 1.049; 1.017 | 0.129; 0.079 | 0.257; 0.230 | 0.921; 0.927 |
| EDR: Nav29 | 0.958; 0.952 | 0.078; 0.064 | 0.105; 0.096 | 0.939; 0.947 |
| EDR: Intelsat | 0.971; 0.963 | 0.061; 0.051 | 0.098; 0.093 | 0.949; 0.956 |
| EDR: Cosmos1263 | 0.994; 0.977 | 0.077; 0.070 | 0.158; 0.151 | 0.947; 0.953 |
| EDR(400-500 km) | 0.970; 0.959 | 0.072; 0.069 | 0.172; 0.164 | 0.946; 0.953 |
| HASDM@GOCE | 1.001; 0.963 | 0.008; 0.009 | 0.113; 0.105 | 0.961; 0.970 |

These densities have been compared to the GOCE-inferred ones, and they are identical except for a scale factor on time scales of an orbit revolution and longer. The GOCE satellite had several, sometimes long, problems causing data gaps, which is why we used the gapless HASDM densities in this study. The mean and standard deviation of the annual density ratio time series, the average RMS and correlation coefficients are listed in Table 3.

The dispersion of the annual time series (10–20%) and the RMS (5–10%) is again significantly smaller, and the correlations higher, with DTM_F30 than with DTM_F10. These improvements may seem small, but in thermosphere modelling progress is made in such relatively small steps.

We have also quantified the modelling improvement thanks to F30 for time scales up to a solar rotation. To that purpose,

density ratios have been binned per ten days for specific years with high variability. Figure 15 displays this for the CHAMP and GRACE satellites, for HASDM. The average of all 10-day density ratios is printed in the figures, as well as the standard deviation of the time series. The standard deviations are again smaller with DTM_F30, which implies again that the solar variability effect on density is also better taken into account.

## 6. Conclusions

The principal conclusion of this study is the superior performance of the 30 cm radio flux over the usual 10.7 cm one when it comes to modelling the response of the upper atmosphere on time scales of days and beyond. The reason for this is the stronger presence in the 30 cm flux of bremsstrahlung that comes from plages and faculae. This bremsstrahlung has a strong counterpart in the UV bands that mostly affects the upper atmosphere. In particular:

- We have built a homogeneous and continuous data set of daily solar radio emissions at five centimetric wavelengths (30, 15, 10.7, 8 and 3.2 cm), extending from 1957 up to today. Such synoptic observations provide considerable added value over the sole F10.7 index because the relative contribution of various emission processes is not the same in them. The data set is available for download from http://projects.pmodwrc.ch/solid/.

- By using a statistical method called blind source separation, for the first time we have been able to empirically decompose the variable component of the radio emissions into three independent contributions. Two of them can be ascribed to gyroresonance emissions and one to bremsstrahlung. The latter accounts for about 10% of the rotational modulation observed at 10.7 cm, and 50% at 30 cm.

- Reconstructions of various solar proxies, and of SSI observations with these five radio fluxes show that the best tradeoff between model complexity and correlation is obtained when a combination of two radio wavelengths is used. Good results are already obtained when using the 10.7 cm and the 30 cm fluxes. The RMS of the reconstructed MgII index, for example, drops by 25% when using both wavelengths instead of the usual 10.7 cm flux only, and by 31% when using all five wavelengths.

- Using the DTM2012 thermosphere model, we find that thermospheric density measurements are systematically better reconstructed with the 30 cm than with the 10.7 cm flux as solar proxy. The RMS of the observed-to-calculated density ratios, which is a standard metric for testing the model fitting capacity, drops on average by 7% when using the 30 cm flux.

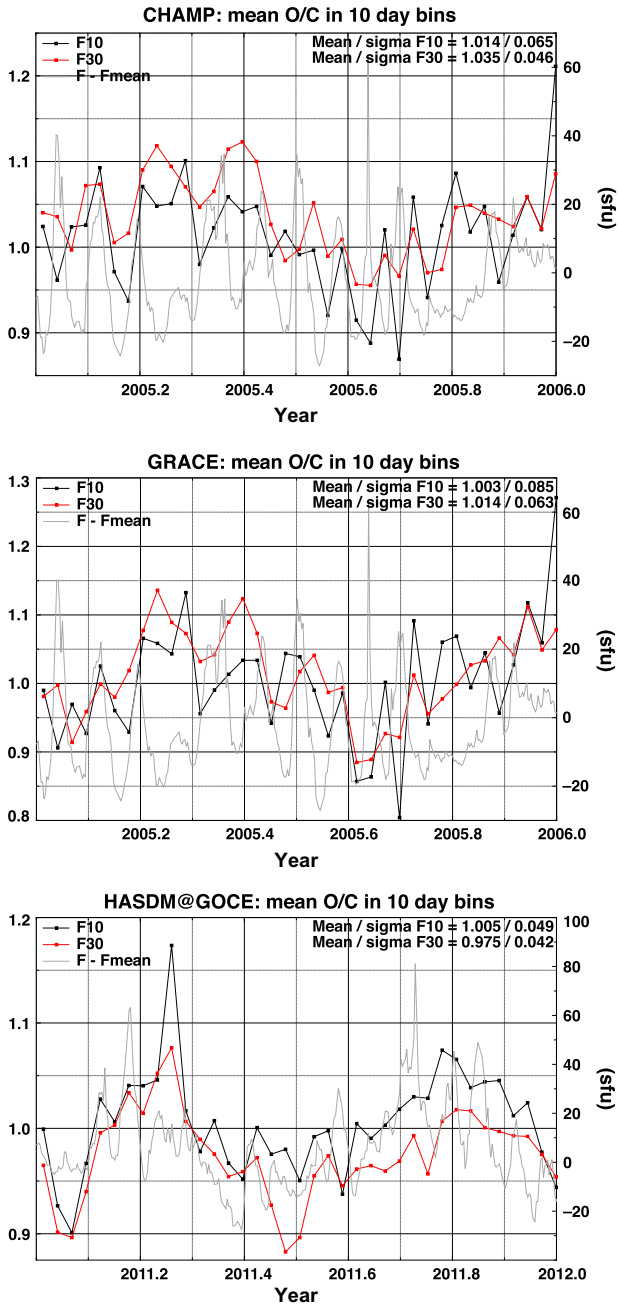

**Figure 15.** Mean *O/C* for 10-day intervals, computed for CHAMP (top), GRACE (middle) and HASDM@GOCE (bottom). Values obtained with DTM2012_F10 are shown in black, and in red for DTM2012_F30. The F10.7 is indicated in grey, in arbitrary units.

The availability of long and well-calibrated records, with over of 50 years of continuous observations available, are additional assets of centimetric radio observations, which have a large potential for space weather and for space climate. New facilities are now starting to monitor the Sun in the centimetric range (e.g. Marqué et al. 2013), which guarantees the continuity of the observations in the next decades.

*Acknowledgements.* This study received funding from the European Community's Seventh Framework Programme (respectively FP7-SPACE-2010-1 and FP7-SPACE-2012-2) under the Grant Agreements 261948 (ATMOP project, www.atmop.eu) and 313188 (SO-LID project, projects.pmodwrc.ch/solid/), and was supported by COST Action ES1005 "TOSCA" (www.tosca-cost.eu). SB is equally supported by CNES/TOSCA The following institutes are acknowledged for providing the data that were used in this study: Laboratory for Atmospheric and Space Physics (Boulder), National Geophysical Data Center (NOAA), the Solar-Terrestrial Centre of Excellence (Brussels), Mount Wilson Observatory and the National Research Council of Canada. This study includes data from the synoptic programme at the 150-Foot Solar Tower of the Mt. Wilson Observatory. The Mt. Wilson 150-Foot Solar Tower is operated by UCLA, with funding from NASA, ONR and NSF, under agreement with the Mt. Wilson Institute. CELIAS/ SEM experiment on the Solar Heliospheric Observatory (SOHO) spacecraft; SOHO is a joint European Space Agency, United States National Aeronautics and Space Administration mission. Finally, we wish to specially thank the two anonymous referees for their constructive and supportive comments that helped improve the manuscript.